\documentclass[11pt,fleqn]{article}
\usepackage{graphicx}

\begin{document}

\title{
{\sf Compatibility of $\alpha_s(M_\tau)$ with $\alpha_s(M_Z)$} }

\author{
T.G. STEELE \\
{\sl Department of Physics and Engineering Physics and}\\
{\sl Saskatchewan Accelerator Laboratory}\\
{\sl University of Saskatchewan}\\
{\sl Saskatoon, Saskatchewan S7N 5C6, Canada.}\\
{V. ELIAS}\\
{\sl Department of Applied Mathematics,} 
{\sl The University of Western Ontario,} \\
{\sl London, Ontario N6A 5B7, Canada. }
}
\maketitle
\begin{abstract}
The current phenomenological determinations of
$\alpha_s\left(M_\tau\right)$ and $\alpha_s\left(M_Z\right)$ 
are shown to be only marginally consistent with the QCD evolution of the strong coupling constant
between $M_Z$ and $M_\tau$.  This  motivates a revised estimate
of $\alpha_s\left(M_\tau\right)$  since the perturbative series
used to extract $\alpha_s\left(M_\tau\right)$
from the $\tau$ hadronic width 
exhibits slow convergence.  Pad\'e summation techniques  provide an estimate 
of these unknown higher-order effects, leading to the revised determination
$\alpha_s\left(M_\tau\right)=0.333\pm 0.030 $. This  value is $10\%$ smaller than 
current estimates, 
improving the compatibility of
phenomenological estimates for $\alpha_s\left(M_\tau\right)$ 
and $\alpha_s\left(M_Z\right)$ with the QCD evolution of the strong coupling constant.
\end{abstract}
The Particle Data Group (PDG) quotes the following values for the
strong coupling constant as determined from  $Z^0$ and $\tau$ decays \cite{pdg}.
\begin{eqnarray}
& &\alpha_s\left(M_\tau\right)=0.370\pm 0.033
\label{alpha_tau_pdg}
\\
& &\alpha_s\left(M_Z\right)=0.118\pm 0.003
\label{alpha_Z_pdg}
\end{eqnarray}
Since these determinations of $\alpha_s$ occur at such widely separated energies, 
the compatibility of these values of $\alpha_s$ with the QCD evolution of the coupling constant
is an important test of both QCD and the phenomenological results used to extract the coupling constant
from the experimental data.  In particular, $\alpha_s\left(M_\tau\right)$ is sufficiently large
that presently unknown terms from higher order perturbation theory could 
substantially alter the
value of $\alpha_s\left(M_\tau\right)$  extracted from the experimental data.
Pad\'e approximant methods provide estimates of the aggregate effect of
(presently unknown) terms from higher-order perturbation theory \cite{pade,samuel2,apap,samuel}.
As shown below, the use of Pad\'e summation
to estimate such terms 
leads to a substantial decrease in the
value of $\alpha_s\left(M_\tau\right)$ extracted from $\tau$ decays, improving the 
compatibility of $\alpha_s\left(M_\tau\right)$ and $\alpha_s\left(M_Z\right)$ with the QCD evolution of the
coupling constant.

The QCD evolution of the coupling constant is governed by the $\beta$ function
which is now known to 4-loop order \cite{beta}.  Using the conventions of
\cite{chetyrkin1}, $a\equiv\frac{\alpha_s}{\pi}$ satisfies the differential equation
\begin{eqnarray}
& &\mu^2\frac{d a}{d \mu^2}=
\beta(a)=
-a^2\sum_{i=0}^\infty \beta_i a^i\quad ,\quad a\equiv\frac{\alpha_s}{\pi}
\label{evolution}
\\
& &\beta_0=\frac{11-\frac{2}{3}n_f}{4}
\quad ,\quad
\beta_1=\frac{102-\frac{38}{3}n_f}{16}\quad ,\quad
\beta_2=\frac{\frac{2857}{2}-\frac{5033}{18}n_f+\frac{325}{54}n_f^2}{64}
\\[5pt]
& &
\beta_3=114.23033-27.133944 n_f+1.5823791 n_f^2-5.8566958\times 10^{-3}n_f^3
\end{eqnarray}
Using the value  $\alpha_s\left(M_z\right)$ as an initial condition, the coupling
constant can be evolved to the desired energy using the differential equation
(\ref{evolution}).  The only subtlety in this approach is the location of flavour thresholds
where the number of effective flavour degrees of freedom $n_f$ change.  In general, matching conditions
must be imposed at these thresholds to relate QCD with   $n_f$ quarks to an effective theory
with $n_f-1$ light quarks and a decoupled heavy quark \cite{rodrigo}.  Using the matching threshold $\mu_{th}$ defined by
$m_q(\mu_{th})=\mu_{th}$, where $m_q$ is the running quark mass, the matching condition to three-loop order is 
\cite{chetyrkin2}
\begin{equation}
a^{(n_f-1)}\left(\mu_{th}\right)=
a^{(n_f)}\left(\mu_{th}\right)\left[1+0.1528 \left[a^{(n_f)}\left(\mu_{th}\right)\right]^2
+\left\{0.9721-0.0847\left(n_f-1\right)\right\}\left[a^{(n_f)}\left(\mu_{th}\right)\right]^3\right]
\label{matching}
\end{equation}
leading to a discontinuity of $\alpha_s$ across the threshold.
Thus to determine the coupling constant at energies between the $c$ quark threshold and the $b$ quark threshold,
the $\beta$ function with $n_f=5$ is used to run $\alpha_s^{(5)}$ from $M_Z$ to $\mu_{th}=m_b(\mu_{th})\equiv m_b$ using
(\ref{alpha_Z_pdg}) as an initial condition.  The matching condition
(\ref{matching}) is then imposed to find the value of $\alpha_s^{(4)}(m_b)$ which is then used as an initial condition
to evolve $\alpha_s$ to lower energies via the $n_f=4$ $\beta$ function.

If $\alpha_s\left(M_Z\right)$ is used as the input value to determine
the QCD prediction of $\alpha_s\left(M_\tau\right)$, then  one might legitimately be concerned about the effect of
(unknown) higher-order terms in the $\beta$ function at lower energies where $\alpha_s$ is larger.
Pad\'e approximations have proven their utility in determining higher-order terms in the $\beta$ function.
For example, using as input the four-loop $\beta$ function in $O(N)$ gauge theory \cite{O(N)},
asymptotic Pad\' e methods
described in Section II of \cite{apap}
are able to
predict the five-loop term to better than $10\%$ of the known five-loop contributions for $N\le 4$ \cite{samuel2,elias}.  
When these same methods are applied to QCD, the following predictions for the unknown five-loop contribution to the $\beta$
function are obtained \cite{elias}.
\begin{eqnarray}
n_f=4: & & \beta_4=83.7563
\label{beta_5_nf=4}
\\
n_f=5: & & \beta_4=134.56
\label{beta_5_nf=5}
\end{eqnarray}
From these predictions, $\beta$ functions containing $[2\vert 2]$ Pad\'e approximants
can be constructed to estimate the sum of all higher-order contributions.
These Pad\'e-summations, whose Maclaurin expansions reproduce $\beta_1,~\beta_2,~\beta_3$ and
and the asymptotic Pad\'e-approximant estimates (\ref{beta_5_nf=4},\ref{beta_5_nf=5}) of
$\beta_4$, are given by:
\begin{eqnarray}
n_f=4: & & \beta(a)=-\frac{25x^2}{12}\left[\frac{1-5.8963 a-4.0110 a^2}{1-7.4363 a+4.3932 a^2}\right]
\label{beta_pade_nf4}
\\[5pt]
n_f=5: & & \beta(a)=-\frac{23x^2}{12}\left[\frac{1-5.9761 a-6.9861 a^2}{1-7.2369 a-0.66390 a^2}\right]
\label{beta_pade_nf5}
\end{eqnarray}
Thus the QCD prediction of $\alpha_s\left(M_\tau\right)$ depends on only two parameters: the initial condition
$\alpha_s\left(M_z\right)$ and the position of the five-flavour threshold defined by $m_b(m_b)=m_b$.  
As will be discussed below, the uncertainty in the Particle Data Group value \cite{pdg} for this threshold
\begin{equation}
m_b(m_b)=(4.3\pm 0.3){\rm GeV}
\label{mb_pdg}
\end{equation}
has a negligible effect on the QCD prediction of $\alpha_s\left(M_\tau\right)$ compared with the
uncertainty in $\alpha_s\left(M_z\right)$ (\ref{alpha_Z_pdg}).

\begin{figure}
\centering
\includegraphics[scale=0.7]{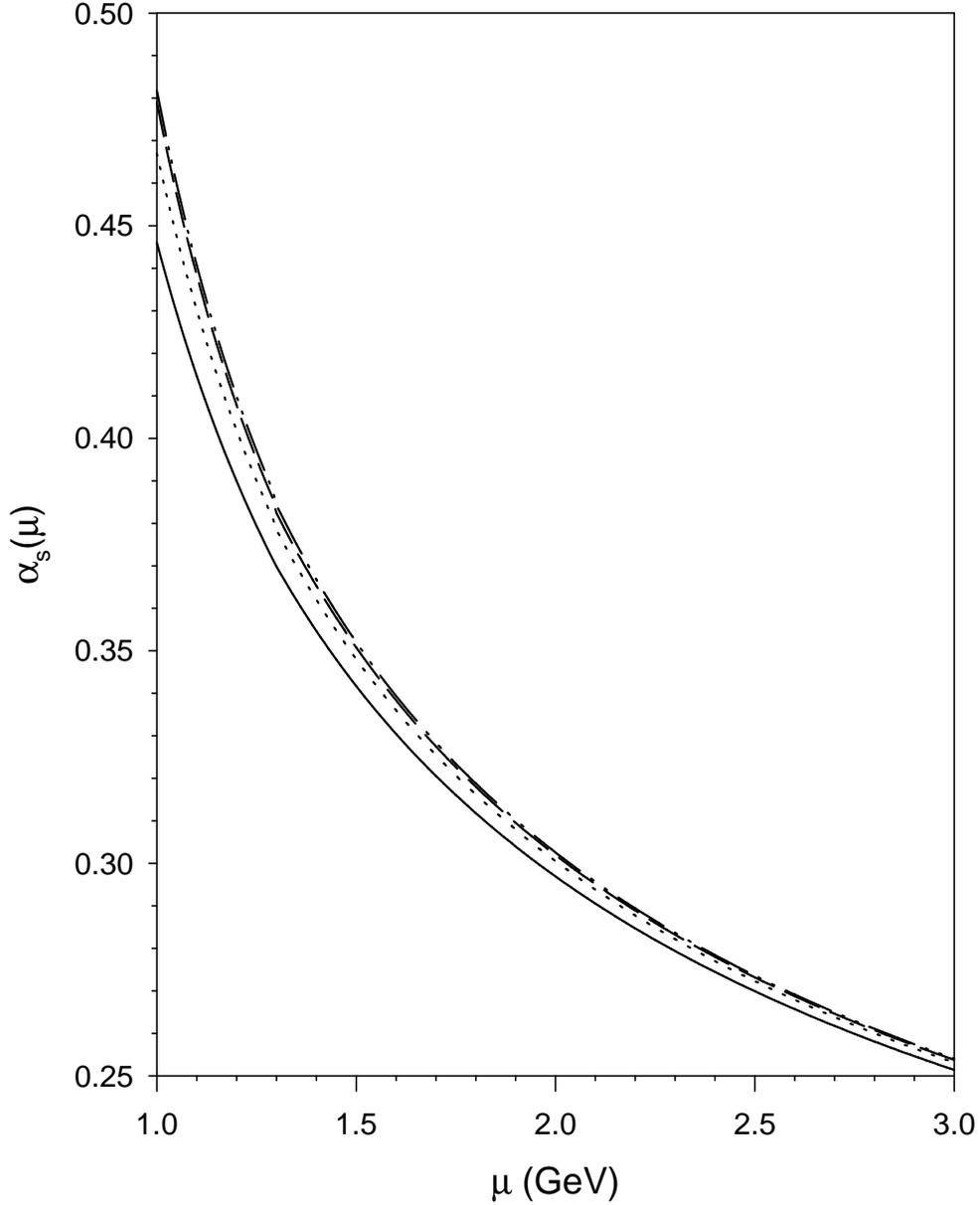}
\caption{
Effect of increasing the order of perturbation theory 
in the QCD evolution of the strong coupling constant using 
$\alpha_s\left(M_Z\right)$ as an initial condition.
Higher-loop terms in the $\beta$ function progressively increase
$\alpha_s$ from the 2-loop order bottom (solid) curve to the
Pad\'e summation top (dashed-dotted) curve, sandwiching the
three- and four-loop curves. 
}
\label{alphafig}
\end{figure}

The compatibility of the experimentally/phenomenologically determined  values $\alpha_s\left(M_Z\right)$ 
and $\alpha_s\left(M_\tau\right)$ with the QCD evolution of the coupling constant can now be studied.
Figure \ref{alphafig}   shows the effect on $\alpha_s(Q)$ of progressive increases in the number of
perturbative terms in the $\beta$ function, culminating with the Pad\'e summation 
(\ref{beta_pade_nf4},\ref{beta_pade_nf5}) for $\beta$.  It is evident that 
the curves for $\alpha_s(Q)$ appear to converge from below to that generated by the Pad\'e summation
of the $\beta$ function, since the gaps between curves of successive order decrease.
Using the input values (\ref{alpha_Z_pdg},\ref{mb_pdg}) for the QCD evolution of $\alpha_s$ down from the $Z^0$ to $\tau$ mass, 
we obtain the following range of values for $\alpha_s\left(M_\tau\right)$ for successive orders of
perturbation theory \cite{elias}:
\begin{eqnarray}
& &{\rm 2-loop} \qquad 0.2910\le\alpha_s\left(M_\tau\right) \le 0.3391
\label{2loop_alpha_mtau}\\
& &{\rm 3-loop} \qquad 0.2944\le\alpha_s\left(M_\tau\right) \le 0.3451
\label{3loop_alpha_mtau}\\
& &{\rm 4-loop} \qquad 0.2957\le\alpha_s\left(M_\tau\right) \le 0.3477
\label{4loop_alpha_mtau}\\
& &{\rm Pade~ summation} \qquad 0.2963\le\alpha_s\left(M_\tau\right) \le 0.3489
\label{pade_alpha_mtau}
\end{eqnarray} 
The dominant effect on the uncertainty in the above originates from $\alpha_s\left(M_Z\right)$---
the effect of the uncertainty in the five-flavour threshold (\ref{mb_pdg}) is inconsequential.
Thus as illustrated in Figure \ref{compatfig}, the  empirical determinations of $\alpha_s$ at $M_Z$ and $M_\tau$ are only marginally consistent with the QCD
evolution of the coupling constant.  This motivates a revised determination of $\alpha_s\left(M_\tau\right)$
using Pad\'e approximation techniques.

The seminal QCD analysis of the $\tau$ hadronic width \cite{braaten}
\begin{equation}
R_\tau\equiv \frac{\Gamma\left(\tau\rightarrow \nu_\tau+{\rm hadrons}\right)}{
\Gamma\left(\tau\rightarrow \nu_\tau+e+\bar\nu_e\right)}
\end{equation}
leads to the following result for $R_\tau$.
\begin{equation}
R_\tau=3\left(\left\vert V_{ud}\right\vert^2+\left\vert V_{us}\right\vert^2\right)S_{EW}\left[
1+\delta'_{EW}+\delta^{(0)}+\sum_{D=2,4,6} \cos^2\theta_c\delta^{(D)}_{ud}+
\sin^2\theta_c\delta^{(D)}_{us}
\right]
\label{R_tau}
\end{equation}
To facilitate comparison of the Pad\'e improvement of this result with previous research
we use values for the parameters in (\ref{R_tau}) identical to \cite{braaten}:
the CKM matrix elements,
\begin{equation}
\left\vert V_{ud}\right\vert=0.9753\quad ,\quad \left\vert V_{us}\right\vert=0.2210
\label{ckm}
\end{equation}
and the Cabibbo angle
\begin{equation}
\sin^2\theta_c=\frac{\left\vert V_{us}\right\vert^2}{\left\vert V_{ud}\right\vert^2+\left\vert V_{us}\right\vert^2}
\end{equation}
The electroweak contributions in (\ref{R_tau}) are
\begin{equation}
S_{EW}=1.0194\quad ,\quad \delta'_{EW}=0.001 \quad .
\label{delta_EW}
\end{equation}

The mass-independent perturbative contribution to $R_\tau$ is the dominant effect: in the $\overline{\rm MS}$ scheme, 
it takes the form 
\begin{equation}
1+\delta^{(0)}=1+a\left(M_\tau\right)+5.2023 \left[a\left(M_\tau\right)\right]^2
+26.366\left[a\left(M_\tau\right)\right]^3 \quad .
\label{delta_0}
\end{equation}
Note that for a characteristic value $\alpha_s\left(M_\tau\right)=0.37$, the contributions
of various orders to $\delta^{(0)}$ are 
\begin{equation}
1+\delta^{(0)}=1+0.118+0.072+0.043
\end{equation}
illustrating the slow convergence of the perturbation series, and hence the
potentially important role of presently unknown terms
from higher-order perturbation theory.

Perturbative quark mass contributions to $R_\tau$ occur for $D=2$ in the summation.  The quantity $\delta^{(2)}_{ij}$, representing
the average of the axial and vector contributions to $R_\tau$, is
\begin{equation}
\delta^{(2)}_{ij}=-8\left[1+\frac{16}{3} a\left(M_\tau\right)\right]\frac{m^2_i\left(M_\tau\right)+
m^2_j\left(M_\tau\right)}{M_\tau^2}
\label{delta_2}
\end{equation}
where $m_i\left(M_\tau\right)$ denotes the running quark mass (of flavour $i$) which has
{\em implicit} dependence on $\alpha_s\left(M_\tau\right)$ through its running behaviour
\begin{eqnarray}
& &m_i(\mu)=\hat m_i\left[2\beta_0 a(\mu)\right]^{\frac{\gamma_0}{2\beta_0}}\,
\left[1+\frac{\beta_1}{\beta_0}\left(-\frac{\gamma_0}{2\beta_0}+\frac{\gamma_1}{2\beta_1}\right) a(\mu)\right]
\label{run_mass}
\\
& &\gamma_0=2\quad ,\quad \gamma_1=\frac{101}{12}-\frac{5}{18}n_f
\end{eqnarray}
To facilitate comparison  of our results with \cite{braaten}, we use identical
values for the RG invariant quark masses $\hat m_i$.
\begin{equation}
\hat m_u=8.7\,{\rm MeV}\quad ,\quad \hat m_d=15.4\,{\rm MeV}\quad ,\quad
\hat m_s=270\,{\rm MeV}
\end{equation}
The total effect of these $D=2$ mass corrections is small in comparison
with $\delta^{(0)}$ since the potentially large $s$ quark contributions are suppressed
by $\sin\theta_c$ in (\ref{R_tau}).

Dimension-four ($D=4$) and higher contributions to $R_\tau$ are dominantly non-perturbative in origin, and involve
the QCD condensates which originate from the operator-product expansion \cite{svz}.
The average of the axial and vector contributions for $D=4$ is
\begin{eqnarray}
\delta^{(4)}_{ij}&=&\frac{11\pi^2 a^2\left(M_\tau\right)}{4M_\tau^4}\langle a GG\rangle+\frac{16\pi^2}{M_\tau^4}
\left(1+\frac{27}{8}a^2\left(M_\tau\right)\right)\langle m_i\bar \psi_i\psi_i+
m_j\bar \psi_j\psi_j\rangle
\nonumber\\
& &-\frac{8\pi^2 a^2\left(M_\tau\right)}{M_\tau^4}\sum_{k=u,d,s}
\langle m_k\bar \psi_k\psi_k\rangle+\frac{24 m_i^2\left(M_\tau\right)m_j^2\left(M_\tau\right)}{M_\tau^4}
\nonumber\\
& &+\left(-\frac{48}{7 a\left(M_\tau\right)}+\frac{22}{7}\right)
\frac{m_i^4\left(M_\tau\right)+m_j^4\left(M_\tau\right)}{M_\tau^4}
\label{delta_4}
\end{eqnarray}
which  has explicit and implicit dependence on $\alpha_s\left(M_\tau\right)$.
Values for the QCD condensates appearing in (\ref{delta_4}), as used in
\cite{braaten}, are
\begin{eqnarray}
& &\langle a GG\rangle=0.02\,{\rm GeV^4}\\
& &\langle m_j \bar\psi_j\psi_j\rangle\equiv -\hat m_j\hat\mu_j^3
\quad ,\quad
\hat\mu_d=\hat\mu_u=189\,{\rm MeV}\quad ,\quad \hat\mu_s=160\,{\rm MeV}
\end{eqnarray}

Dimension-six contributions are the dominant non-perturbative effect.  After use of 
the effective phenomenological equivalent of the 
vacuum saturation hypothesis \cite{svz} parametrized by the (near) RG-invariant
quantity, 
\begin{equation}
\rho\alpha_s\langle\bar\psi\psi\rangle^2=3.8\times 10^{-4}{\rm GeV^6}
\end{equation}
the average of the axial and vector $D=6$ contributions is \cite{braaten}
\begin{equation}
\delta^{(6)}_{ij}=-\frac{512\pi^3}{27M_\tau^6}\rho\alpha_s\langle\bar\psi\psi\rangle^2
=-0.007
\label{delta_6}
\end{equation}
Dimension-eight and higher dimensional non-perturbative contributions to $R_\tau$ are 
negligible \cite{braaten}.

Pad\'e approximation techniques can now be applied to the determination of
$\alpha_s\left(M_\tau\right)$ from an experimentally determined $R_\tau$.
Asymptotic Pad\'e approximation (APAP) techniques use convergence properties of the perturbative
expansion to provide improved estimates of the fifth order term in a perturbation series
given knowledge of the fourth-order series \cite{apap}.  
Explicit field-theoretical tests of the  APAP
approach where the fifth order terms are known lead to extremely accurate results.
For example, the $N=1$ limit of the five-loop $\beta$ function for $O(N)$ gauge
theory has been calculated explicitly \cite{O(N)}. Using the APAP algorithm
described in Section II of \cite{apap}, it is possible to make an APAP estimate of
the known five-loop term $\beta_5$ based upon the contributions from the previous
four terms $\beta_1$--$\beta_4$.  This estimate has been found to be well within $0.2\%$
of the answer obtained by direct calculation \cite{O(N)}.

Applying the same APAP algorithm to the perturbation series (\ref{delta_0})
leads to the following
prediction for the $\left[a\left(M_\tau\right)\right]^4$ perturbative contribution:
\begin{equation}
1+\delta^{(0)}=1+a\left(M_\tau\right)+5.2023 \left[a\left(M_\tau\right)\right]^2
+26.366\left[a\left(M_\tau\right)\right]^3
+132.44\left[a\left(M_\tau\right)\right]^4
\label{trunc}
\end{equation}
It is significant to note that this prediction is very close to the {\em maximum} estimated size 
of the fourth order effect used to determine the theoretical uncertainty in \cite{braaten}, indicating
an underestimate of the higher order effects in previous work.
The APAP result in (\ref{trunc}) should be compared with that obtained previously  from (non-asymptotic) Pad\'e methods
\cite{samuel}
\begin{equation}
1+\delta^{(0)}=1+a\left(M_\tau\right)+5.2023 \left[a\left(M_\tau\right)\right]^2
+26.366\left[a\left(M_\tau\right)\right]^3
+109.2\left[a\left(M_\tau\right)\right]^4
\label{mark}
\end{equation}
The coefficient of the $a^4$ term is $20\%$ smaller in the non-asymptotic prediction.

To determine  the effect of the APAP prediction (\ref{trunc}) on $\delta^{(0)}$ we again consider
a characteristic value  $\alpha_s\left(M_\tau\right)$
\begin{equation}
\delta^{(0)}=0.118+0.072+0.043+0.025
\end{equation}
The convergence of this series is still quite slow: the 
${\cal O}(a^4)$ contribution is $20\%$ of the leading perturbative contribution.
Such slow convergence indicates that the further higher-order terms
could have a significant effect on $\delta^{(0)}$.  A Pad\'e summation, in this case a
$[2\vert 2]$ Pad\'e approximant, provides a means of estimating the total effect of
higher order terms in the perturbation series. Using the APAP determination in (\ref{trunc})
leads to the following $[2\vert 2]$ Pad\'e approximation for 
$\delta^{(0)}$.
\begin{equation}
1+\delta^{(0)}=\frac{1-6.5483 a\left(M_\tau\right)+10.5030 \left[a\left(M_\tau\right)\right]^2}{1-7.5483 a\left(M_\tau\right)+12.8514 \left[a\left(M_\tau\right)\right]^2}
\label{pade}
\end{equation}

\begin{figure}
\centering
\includegraphics[scale=0.7]{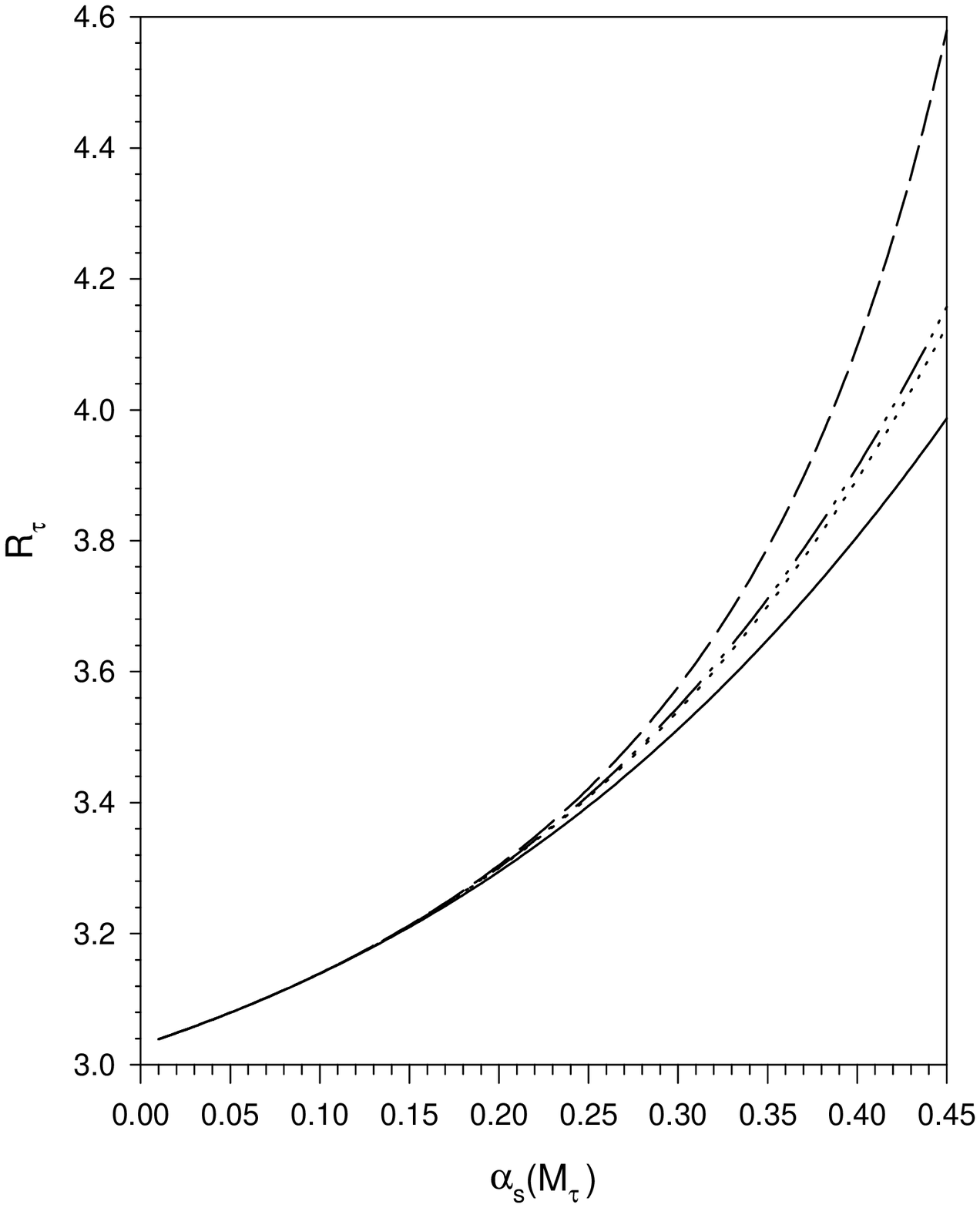}
\caption{$R_\tau$ as a function of $\alpha_s\left(M_\tau\right)$ using 
the four different treatments of $\delta^{(0)}$.  The solid curve uses the
perturbative expression (\protect\ref{delta_0}),
the dotted curve uses the non-asymptotic Pad\'e
(\protect\ref{mark}), the dashed-dotted curve uses the APAP
expression (\protect\ref{trunc}),
and the dashed curve uses  the Pad\'e summation
(\protect\ref{pade}).
}
\label{r_tau_fig}
\end{figure}

Effects of the Pad\'e improvement of $\delta^{(0)}$ on the $\alpha_s\left(M_\tau\right)$ dependence of $R_\tau$, 
and hence the determination of $\alpha_s\left(M_\tau\right)$, can now be investigated
in both the truncated form
(\ref{trunc}) or $[2|2]$ summation form (\ref{pade}).  Figure \ref{r_tau_fig}
compares the
dependence of $R_\tau$ on $\alpha_s\left(M_\tau\right)$  
using the full $D=2,4,6$ contributions [equations (\ref{delta_2},\ref{delta_4},\ref{delta_6})
and electroweak effects (\ref{delta_EW}) for the four different scenarios for 
$\delta^{(0)}$: the known perturbation series (\ref{delta_0}),
the non-asymptotic Pad\'e expression (\ref{mark}),
the asymptotic Pad\'e (APAP) expression (\ref{trunc}), and the
Pad\'e summation (\ref{pade}).  It is evident that 
at the PDG value $a\left(M_\tau\right)=0.370/\pi=0.117$, 
the Pad\'e effects lead to a significant {\em increase} in $R_\tau$.  In comparison with 
the perturbative result, the size of the $R_\tau$ enhancement
obtained from the $[2\vert 2]$ Pad\'e summation (\ref{pade}) is roughly twice the
enhancement obtained directly from (\ref{trunc},\ref{mark}), indicating the significance of the
higher-order effects estimated in the $[2|2]$ Pad\'e summation.
To facilitate comparison with Table 4 in \cite{braaten} and for future determination of 
$\alpha_s\left(M_\tau\right)$, Table \ref{R_tau_table} contains $R_\tau$ for 
selected values of $\alpha_s\left(M_\tau\right)$ in the four scenarios
\footnote{The agreement of the ``perturbative'' column of the Table with \cite{braaten}
provides a consistency check on our calculations.}.  

\begin{figure}
\centering
\includegraphics[scale=0.7]{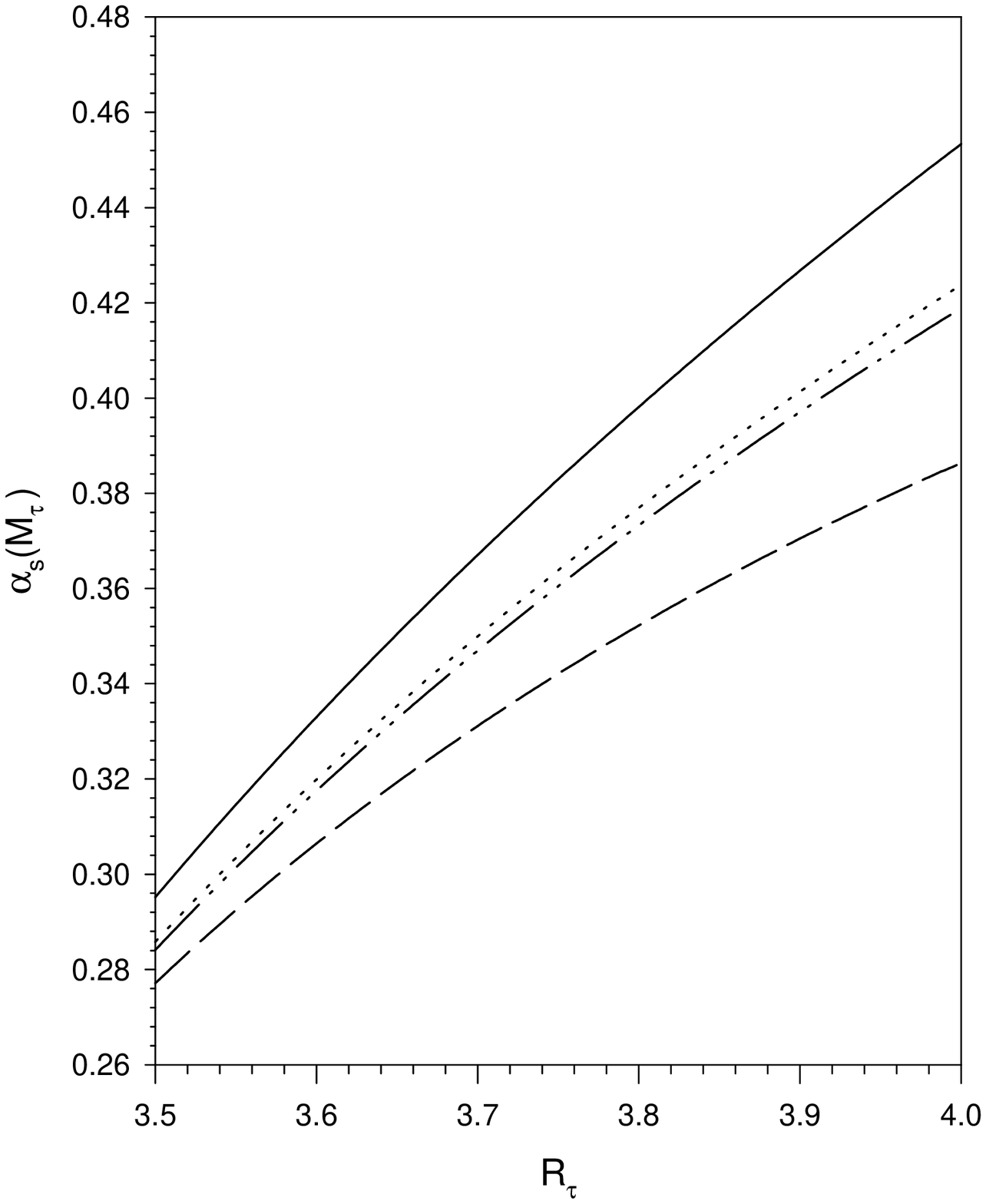}
\caption{$\alpha_s\left(M_\tau\right)$ as a function of $R_\tau$ using
the four different treatments of $\delta^{(0)}$.  The solid curve uses the
perturbative expression (\protect\ref{delta_0}),
the dotted curve uses the non-asymptotic Pad\'e
(\protect\ref{mark}), the dashed-dotted curve uses the APAP
expression (\protect\ref{trunc}),
and the dashed curve uses  the Pad\'e summation
(\protect\ref{pade}).
}
\label{alpha_mtau_fig}
\end{figure}

The Pad\'e enhancement of $R_\tau$ implies that the value of 
$\alpha_s\left(M_\tau\right)$ extracted from the experimental measurement of 
$R_\tau$ will  {\em decrease}.  This is illustrated in Figure \ref{alpha_mtau_fig} which inverts the
relation between $R_\tau$ and $\alpha_s\left(M_\tau\right)$.  
Again, in comparison with 
the perturbative result, the decrease in $\alpha_s\left(M_\tau\right)$ 
following from the Pad\'e summation (\ref{pade}) is roughly twice the decrease one obtains 
from the Pad\'e estimate (\ref{trunc},\ref{mark}) of just the next-order contribution to
$\delta^{(0)}$.
 For the perturbatively-extracted PDG value
$\alpha_s\left(M_\tau\right)=0.37$, the Pad\'e summation would lead to a decrease 
of approximately $10\%$ to $\alpha_s\left(M_\tau\right)=0.33$.

In conclusion, Pad\'e summation methods indicate that the PDG central 
value $\alpha_s\left(M_\tau\right)=0.37$ 
should be brought down by approximately $10\%$ to include the estimated effect 
of higher order QCD corrections.
Assuming the same relative uncertainty characterizing the PDG estimate,  
we find that
\begin{equation}
\alpha_s\left(M_\tau\right)=0.333\pm 0.030 
\end{equation}
a result in much closer agreement with the ranges 
(\ref{4loop_alpha_mtau},\ref{pade_alpha_mtau}) devolving from $\alpha_s\left(M_Z\right)$
via four-loop and Pad\'e-improved four-loop QCD $\beta$ functions.
Thus, Pad\'e methods are seen to improve the compatibility of
$\alpha_s\left(M_\tau\right)$ 
and $\alpha_s\left(M_Z\right)$ with the coupling constant evolution 
anticipated from QCD.

\noindent
{\bf Acknowledgements:}  The authors are grateful for the financial support
of the Natural Sciences and Engineering Research Council of Canada (NSERC).

\begin{table}
\centering
\begin{tabular}{||c|c|c|c|c||}\hline\hline
$\alpha_s\left(M_\tau\right)$ & \multicolumn{4}{c||}{$R_\tau$}\\
 & Perturbative  & Nonasymptotic Pad\'e & APAP & Pad\'e Summation \\\hline\hline
       .20& 3.295& 3.300& 3.301& 3.305\\ \hline
       .21& 3.314& 3.320& 3.322& 3.326  \\ \hline
       .22& 3.333& 3.341& 3.343& 3.348  \\ \hline
       .23& 3.353& 3.362& 3.364& 3.371  \\ \hline
       .24& 3.374& 3.385& 3.387& 3.396  \\ \hline
       .25& 3.395& 3.408& 3.411& 3.422  \\ \hline
       .26& 3.417& 3.433& 3.436& 3.449   \\ \hline
       .27& 3.440& 3.458& 3.462& 3.478   \\ \hline
       .28& 3.463& 3.484& 3.489& 3.509   \\ \hline
       .29& 3.487& 3.511& 3.517& 3.542   \\ \hline
       .30& 3.512& 3.540& 3.546& 3.576   \\ \hline
       .31& 3.538& 3.570& 3.576& 3.613   \\ \hline
       .32& 3.564& 3.600& 3.608& 3.653   \\ \hline
       .33& 3.592& 3.632& 3.641& 3.695   \\ \hline
       .34& 3.620& 3.666& 3.675& 3.740   \\ \hline
       .35& 3.649& 3.700& 3.711& 3.789   \\ \hline
       .36& 3.678& 3.736& 3.748& 3.841   \\ \hline
       .37& 3.709& 3.773& 3.787& 3.897   \\ \hline
       .38& 3.741& 3.812& 3.827& 3.958   \\ \hline
       .39& 3.773& 3.852& 3.869& 4.024   \\ \hline
       .40& 3.806& 3.894& 3.913& 4.097   \\ \hline
       .41& 3.841& 3.937& 3.958& 4.175   \\ \hline
       .42& 3.876& 3.982& 4.005& 4.262   \\ \hline
       .43& 3.912& 4.029& 4.054& 4.357   \\ \hline
       .44& 3.949& 4.077& 4.105& 4.469   \\ \hline\hline
\end{tabular}
\caption{
Values of $R_\tau$ for selected $\alpha_s\left(M_\tau\right)$.
The columns differ only in the treatment of $\delta^{(0)}$ used in $R_\tau$:
the ``perturbative'' column uses
(\protect\ref{delta_0}),
the ``Pad\'e summation'' column uses
(\protect\ref{pade}),
the ``APAP'' column uses
(\protect\ref{trunc}),
and the ``Nonasymptotic Pad\'e'' uses
(\protect\ref{mark}).
}
\label{R_tau_table}
\end{table}

\clearpage

\clearpage
\setlength{\unitlength}{0.5cm}
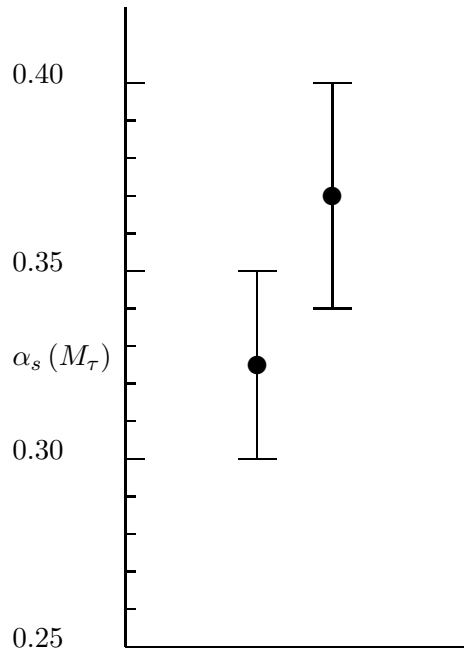
\begin{figure}
\centering
\begin{picture}(15,19)
\put(4,1){\line(0,1){17}}
\put(4,1){\line(1,0){9}}
\multiput(4,1)(0,5){4}{\line(1,0){0.5}}
\multiput(4,1)(0,1){16}{\line(1,0){0.25}}
\put(7.5,8.5){\circle*{0.5}}
\put(9.5,13){\circle*{0.5}}
\put(7.5,6){\line(0,1){5}}
\put(9.5,10){\line(0,1){6}}
\put(7,6){\line(1,0){1}}
\put(7,11){\line(1,0){1}}
\put(9,10){\line(1,0){1}}
\put(9,16){\line(1,0){1}}
\put(1,1){0.25}
\put(1,6){0.30}
\put(1,8.5){$\alpha_s\left(M_\tau\right)$}
\put(1,11){0.35}
\put(1,16){0.40}
\end{picture}
\caption{Comparison of $\alpha_s\left(M_\tau\right)$ extracted from experiment 
(right data point) and $\alpha_s\left(M_\tau\right)$ determined by  
QCD evolution of $\alpha_s$ from $M_Z$ to $M_\tau$ (left data point). }
\label{compatfig}
\end{figure}

\clearpage

\clearpage


\clearpage
\begin{thebibliography}{99}

\bibitem{pdg} Particle data Group, R.M. Barnett {\it et al} Phys. Rev.  {\bf D54} (1996) 1.

\bibitem{pade} M.A.~Samuel, G.~Li, E.~Steinfelds, Phys. Rev. {\bf D48} (1993) 869;
M.A.~Samuel, G.~Li, E.~Steinfelds, Phys. Rev. {\bf E51} (1995) 3911.

\bibitem{samuel2} J.~Ellis, M.~Karliner, M.A.~Samuel,
Phys. Lett. {\bf B400} (1997) 176.

\bibitem{apap} J.~Ellis, I.~Jack, D.R.T.~Jones, M.~Karliner, M.A.~Samuel,
Phys. Rev {\bf D57} (1998) 2665.

\bibitem{samuel} M.A.~Samuel, J.~Ellis,~M.~Karliner, Phys. Rev. Lett. {\bf 74} (1995) 3911.

\bibitem{beta} T.~van~Ritbergen, J.A.M.~Vermaseren, S.A.~Larin, Phys. Lett. {\bf B400} (1997) 379.

\bibitem{chetyrkin1} K.G. Chetyrkin, Phys. Lett. {\bf B404} (1997) 161.

\bibitem{rodrigo} G. Rodrigo, A.~Santamaria, Phys. Lett. {\bf B313} (1993) 441.

\bibitem{chetyrkin2} K.G. Chetyrkin, B.A.~Kniehl, M.~Steinhauser, hep-ph/9708255 v2

\bibitem{O(N)} H. Kleinart, J.~Neu.~V.~ Schulte-Frohlinde, K.G.~Chetyrkin, S.A.~Larin,
Phys. Lett. {\bf B272} (1991) 36, Phys. Lett. {\bf B319} (1993) 545 Erratum. 

\bibitem{elias} V. Elias, T.G.~Steele, F.~Chistie, R.~Migneron, K.~Sprague,
hep-ph/9806324 ~.

\bibitem{braaten} E. Braaten, S.~Narison, A.~Pich, Nucl. Phys. {\bf B373} (1992) 581.

\bibitem{svz} M.A.~Shifman, A.I.~Vainshtein, V.I.~Zakharov,  Nucl. Phys. {\bf B147} (1979) 385.



\end{thebibliography}
\end{document}